\documentclass[10pt]{article}
\usepackage{amsmath,amssymb}
\usepackage{graphicx,booktabs}
\newcommand{\abs}[1]{\lvert #1\rvert}
\newcommand{\avg}[1]{\langle #1\rangle}

\usepackage[labelfont=bf,labelsep=period,justification=raggedright]{caption}

\makeatletter
\renewcommand{\@biblabel}[1]{\quad#1.}
\makeatother

\date{}

\begin{document}

\begin{flushleft}
{\Large\bfseries Emergence of scale-free leadership structure in social recommender systems}\\
Tao Zhou$^{1,2,3,\ast}$, Mat\'u\v{s} Medo$^2$, Giulio Cimini$^{2}$, Zi-Ke Zhang$^{1,2}$, Yi-Cheng Zhang$^{1,2}$\\
\bf{1} Web Sciences Center, University of Electronic Science and Technology of China, Chengdu 610054, People's Republic of China\\
\bf{2} Department of Physics, University of Fribourg, Chemin du Mus\'ee 3, Fribourg 1700, Switzerland\\
\bf{3} Department of Modern Physics, University of Science and Technology of China, Hefei 230026, People's Republic of China\\
$\ast$ E-mail: zhutou@ustc.edu
\end{flushleft}

\section*{Abstract}
The study of the organization of social networks is important
for understanding of opinion formation, rumor spreading, and the
emergence of trends and fashion. This paper reports empirical
analysis of networks extracted from four leading sites with
social functionality (Delicious, Flickr, Twitter and YouTube)
and shows that they all display a scale-free leadership
structure. To reproduce this feature, we propose an adaptive
network model driven by social recommending. Artificial
agent-based simulations of this model highlight a ``good get
richer'' mechanism where users with broad interests and good
judgments are likely to become popular leaders for the others.
Simulations also indicate that the studied social recommendation
mechanism can gradually improve the user experience by adapting
to tastes of its users. Finally we outline implications for real
online resource-sharing systems.
% the previous version of the abstract:
% Uncovering the organization of the social leadership networks is
% significant for the understanding of the rumor spreading and the
% formation of public opinion and fashion. This Letter reports the
% empirical analysis on the leader-follower network extracted from the
% world largest social bookmarking site, from which a scale-free
% leadership structure is observed. Accordingly, we propose an
% adaptive network model driven by social recommending, which could
% reproduce the empirical observations. This model highlights the
% ``good gets richer" mechanism where a user with wider scope of
% interests and better judgment will probably become a leader.
% Further numerical simulations indicate that the proposed social
% recommending mechanism could largely improve the user experience.
% Finally, we outline the implications of this work to the real-world
% online resource-sharing systems.

\section*{Introduction}
Social network analysis has become a joint focus of many
branches of science \cite{Watts2007,Borgatti2009}. Various
social networks have been systematically investigated, such as
friendship, membership and co-authorship networks. In this work
we focus on the so-called leadership networks which capture
how people copy actions or receive information of others.
Although they play a significant role in formation and
propagation of social opinions, leadership networks have
received considerably less attention than other social
networks---possibly because of the lack of empirical data.
Recently, some researchers reported the emergence of scale-free
leadership structures from initially homogeneous interaction
networks in evolutionary games, such as the \emph{minority game}
\cite{Anghel2004,Lo2005,Lee2006}, the \emph{ultimatum game}
\cite{Savarimuthu2008} and the prisoner's dilemma game \cite{Szolnoki2008,Szolnoki2009a,Szolnoki2009b,Poncela2008}, where agent $i$ is considered to be led by agent $j$ if $i$ has adopted $j$'s strategy. Since it is hard
to automatically extract \emph{who follows whom} from records of
economic activities, up to now no empirical evidence has been
reported to either support or challenge these findings for
economic systems. On the other hand, web activity data give us
the possibility to study leadership structures in the process of
information propagation. In this paper, we report both empirical
evidence and a theoretical model for the emergence of scale-free
leadership networks in online societies. Furthermore, we discuss
which user characteristics are important for becoming a leader.

Beyond providing a mechanism leading to scale-free leadership
structures, this work has potential significance for solving the
information overload problem created by the unceasingly growing
amount of easily available information. \emph{Recommender
systems} provide a solution to this problem by analyzing users'
profiles and past preferences and using them for automated
recommendation of relevant items to individual users
\cite{Resnick1997}. The majority of current recommender systems
use a centralized approach where all data is stored and analyzed
at one place. Typical algorithms include collaborative filtering
\cite{Herlocker2004,Schafer2007}, matrix decomposition
\cite{Maslov2001,gravity2007,Ren2008}, and spreading processes
\cite{Zhang2007,Zhou2007,Zhou2010}. However, this paradigm is challenged
by the findings that social influence often plays a more
important role than similarity of past activities
\cite{Ziegler2004,Bonhard2006} and recommendations made by a
system are preferred less than those coming from our friends
\cite{Sinha2001,Huang2010}. In response, \emph{social recommendation} has
become a candidate for the next recommendation paradigm
\cite{Golbeck2008}. \emph{Social recommender systems} can be
designed (i) in a passive way where a user selects other users
as information sources and can import URLs or subscribe blog
articles from them (as in \emph{delicious.com} and
\emph{blogger.com}) \cite{Hammond2005} or (ii) in an active way
where each user can recommend items to other users who have
accepted him as information source (as in \emph{douban.com} and
\emph{twitter.com}) \cite{Chen2008}. While very different from
the user's point of view, these two ways are similar in how
information favored by one user spreads to the user's followers,
followers' followers, and so on \cite{Medo2009,Cimini2011,Wei2011}. This
process is similar to the well-studied epidemic spreading on
networks \cite{Pastor2003,Zhou2006}. The model proposed and
investigated here mimics information spreading process
in adaptive social networks. We test its efficiency in filtering
out the low-quality and irrelevant information and show that
this distributed social recommender model can enhance the user
experience.

\section*{Empirical Results}
The studied bookmarking data was obtained by crawling the
publicly-available data from the social bookmarking website
\emph{delicious.com} \cite{ZhangZK2010}. The resulting network consists of 392\,251
users and 1\,686\,131 directed links. We say that user $i$ is a
follower of user $j$ (or, equivalently, $j$ is a leader of $i$)
if $i$ has imported some of $j$'s bookmarks. In this way, a
directed social network of users is constructed where each link
represents a leader-follower relationship. We define the
direction of each link as
$\textit{leader}\rightarrow\textit{follower}$ and thus the
out-degree of a user (i.e., the number of user's followers) can
be used to quantify the person's leadership strength.
To obtain a solid understanding of the leadership structure, we
study data from three other social sites containing this kind
of structure: \emph{flickr.com}, \emph{twitter.com} and
\emph{youtube.com}. These data sets were provided upon request
by \cite{Mislove2007} for \emph{flickr.com} and
\emph{youtube.com} and by \cite{Kwak2010} for
\emph{twitter.com}. In the first two cases, user $i$ follows
user $j$ if $i$ has asked user $j$ for friendship and user $j$
accepted this invitation. In the case of Twitter data, users can
explicitly follow other users, who will in turn push messages to
them.

Table 1 summarizes basic statistics of the studied leadership
networks and results of power-law fits of their out-degree
distributions based on the standard maximum likelihood
estimation \cite{Goldstein2004,Clauset2009}. The out-degree
distributions themselves are shown in Fig.~1 together with their
power-law fits in the range $[x_{\mathrm{min}},\infty)$
(according to \cite{Clauset2009}, the optimal value of
$x_{\mathrm{min}}$ is the one yielding the minimal value of the
Kolmogorov-Smirnov statistic).

\section*{Model}
The modeled system consists of $N$ users, each of whom has $M$
information sources (i.e., $M$ leaders). Nodes of
the corresponding directed network are hence of identical
in-degree $M$. The out-degree can be used to quantify the node's
leadership status (see also more complicated measures based on
PageRank \cite{Radicchi2009,Lee2010} or LeaderRank \cite{Lu2011}
algorithms). At each time step, a randomly selected user posts
an item (this generic term stands for an URL, a news, a blog
article, a picture, a video, or any other shared content). This
item is automatically considered to be approved (liked) by this
user and spreads to all user's followers who consequently judge
this item. If a follower approves the item, it spreads farther
to the follower's followers. If the item is disapproved, it does
not spread further from this disapproving node (though, it may
continue to spread from some other nodes which approve it). Note that,
in each time step, one piece of news is introduced and spreads through the whole system
depending on approvals/disapprovals of users. This "fast user evaluations" mechanism is just a skill for
the implementation of simulation, which obviously has no effect on the essential feature of the dynamics.

In the model, leaders are evaluated by their followers on the
basis of how the followers appreciate recommendations coming
from them. In particular, the similarity of evaluations $s_{ij}$
is computed for each leader-follower pair. If user $i$ receives
an item from user $j$ and approves it, the similarity score is
updated as $s_{ij}\leftarrow (1-1/n_{ij})s_{ij}+1/n_{ij}$ while
when this item is disapproved by user $i$,
$s_{ij}\leftarrow (1-1/n_{ij})s_{ij}$. Here $n_{ij}$ denotes the
cumulative number of items that $i$ has received from $j$. This
form ensures that contribution of one incoming item to the
similarity value is inversely proportional to the total number
of items transferred through the corresponding channel. Each
user is initially given $M$ randomly selected leaders whose
similarity values are set to $0.5$. It is easy to prove that the
aforementioned formulas lead to $s_{ij}=a_{ij}/n_{ij}$ where
$a_{ij}$ denotes the number of items received from $j$ and
approved by $i$.

To allow for a gradual evolution of the leader-follower network,
each user updates their leaders after every $T$ evaluated items.
We adopt a simple approach in which the worst-performing leader
(the one with the lowest similarity value) of user $i$ is
dropped and replaced by a randomly selected user $j$ (given $j$
is not among the given user's leaders yet). Similarity of this
new leader is set to $s_{ij}=0.5$ and the number of transferred
items to $n_{ij}=0$, independently of whether $j$ has been $i$'s
leader sometimes before or not. Note that this updating is very
economic as it requires no computation and no centralized data
storage (compared with the expensive network optimization
techniques studied in \cite{Medo2009,Cimini2011}). Yet it
ensures that the system evolves in a self-organized way and
gradually adapts to tastes of its users.

To test the described recommendation algorithm, we introduce a
simple agent-based model. The cornerstone of this model is how
to cast evaluations of items by users. We adopt the approach similar to
\cite{Medo2009} where users and items are
described by $D$-dimensional taste and attribute vectors,
respectively. While elements of the user taste vectors
$\mathbf{u}_i$ are randomly set to either $0$ or $1$ with equal
probabilities, elements of the item attribute vectors
$\mathbf{v}_{\alpha}$ are independently drawn from the uniform
distribution $\mathcal{U}(-1,1)$. Note that for clarity we use
Latin and Greek letters for user- and item-related indices,
respectively. Opinion of user $i$ about item $\alpha$ is modeled
as $r_{i\alpha}=\mathbf{u}_i\cdot\mathbf{v}_{\alpha}/D+
\varepsilon \eta_i$ where $\varepsilon$ is a random variable
drawn from the uniform distribution $\mathcal{U}(-1,1)$ and
$\eta_i$ represents the evaluation noise magnitude of user $i$
(the lower the $\eta_i$, the better the judgment, and vice
versa). In this way, opinion of a user about an item is of a
high value if this user's taste vector highly overlaps with the
news's attribute vector. Values $\eta_i$ are drawn from the
uniform distribution $\mathcal{U}(0,0.5)$ and stay fixed during
the simulation. If $r_{i\alpha}$ is larger than a certain
threshold $R_c$, user $i$ approves item $\alpha$. At every time
step, after user $i$ has been randomly selected to post item
$\alpha$, items with random attributes are generated until one
is approved by this user (i.e., it satisfies the approval
condition $r_{i\alpha}>R_c$). Spreading starts of this item then
starts by pushing it to all followers of user $i$.
%%% activity probability and item submission probability probably
%%% need to be introduced. if that's true, their values need to
%%% be specified in the caption of Figure 2

This agent-based vector model has a simple intuitive
interpretation. Respective item's attributes, ranging from $-1$
to $+1$, represent item's quality in various aspects (the
higher, the better) as well as item's topic (e.g., if it
concerns sport or politics or something else). Respective user's
tastes, ranging from $0$ to $+1$, represent user's sensitivity
to different item attributes. A user whose taste vector mostly
consists of ones is sensitive to all attributes and hence can
judge items well. By contrast, a user whose taste vector mostly
consists of zeros is ignorant to most aspects and can be
satisfied with items that would be judged badly by most users.

\section*{Scale-Free Leadership Structure}
The threshold $R_c$ determines the average spreading range of
items (i.e., their average number of readers $\avg{\Sigma}$).
Although the approval thresholds could differ from one user to
another, for simplicity we set them all identical. As shown in
the lower-left inset of Fig.~2, $\avg{\Sigma}$ decreases quickly
as $R_c$ grows and approaches its lower bound when
$R_c\gtrsim0.35$ (each item is evaluated at least by the user
who submitted it and all followers of this user, hence this
lower bound equals $M+1$). We set $R_c=0.2$ to achieve
$N\gg\avg{\Sigma}\gg 1$. The upper-right inset of Fig.~2 shows
the initial out-degree distributions which are naturally simple
Poisson distributions peaked at $M$. After a certain period of
system's evolution (Fig.~2 displays the results after $10^6$
time steps), a scale-free leadership structure is created with
the scaling exponent $\alpha\approx 1.63$.

Scale-free networks are observed in very diverse systems
\cite{Caldarelli2007} which indicates the existence of distinct
mechanisms of their emergence~\cite{Mitzenmacher2003}. While the
majority of evolving network models are directly or implicitly
inspired by the ``rich get richer'' phenomenon
\cite{Merton1968,Egghe1995,Barabasi1999}, there are plenty of other possible
mechanisms such as the optimal design \cite{Valverde2002},
Hamiltonian dynamics \cite{Baiesi2003}, merging and regeneration
\cite{Kim2005} and stability constraints \cite{Perotti2009}. The
mechanism leading to scale-free structures in our model is
different as it is based on a spreading mechanism in a social
network and user heterogeneity. To uncover which factors make a
popular leader, we characterize user $i$ by the quality of
evaluations and the scope of interests. The former is measured
by the noise level $\eta_i$ and the latter by the coverage
$\abs{\mathbf{v}_i}$ which we define as the sum of the taste
vector's elements (which in our case is equal to the number of
ones in $\mathbf{v}_i$). In Fig.~3, we report how the scope of
interests and quality of evaluations affect the number of
followers. As explained before, users with high
$\abs{\mathbf{v}}$ can better reveal intrinsic quality of items
and hence they are likely to approve items with many positive
entries in their attribute vectors---they are good filters of
the content. If a user cannot find enough taste-mates (users
with similar taste vectors), users who filter well can be used
instead. Therefore, in accordance with the dependencies shown in
Figs.~3a and 3c, users with high coverage usually have large
numbers of followers. The role of quality of evaluations is more
complicated. As shown in Fig.~3d, it is clear that popular
leaders have small $\eta$. However, an accurate user may have a
low popularity (see Fig.~3b: the average out-degree of accurate
users is only slightly higher than that of inaccurate users)
because however accurate user $i$ is, if his scope is not broad
enough, the number of users with similar taste is limited.

We also studied the case where some users are more active than
the others (they post and evaluate items more frequently). In
the early stage, the active users have good chance to become
popular leaders but in the long term, the popularity difference
between active and normal users vanishes. This suggests that it
is indeed the intrinsic personal features---scope of interests
and quality of evaluations---what plays the crucial role in
determining a user's position in the social leadership network.
We further investigated cases where (i) users have identical
noise levels, (ii) users have identical coverage, (iii) users
are all alike. In all these cases, the resulting out-degree
distributions are considerably narrower than those reported in
Fig.~2. Together with big standard deviations observed in
Figs.~3a and 3b for large $\abs{\mathbf{v}}$ and small $\eta$,
we can conclude that each of the qualities alone is not enough:
popular leaders are those who have both broad scope and little
randomness in their evaluations. This is similar to the ``good
get richer'' mechanism proposed in the study of complex networks
\cite{Caldarelli2002,Garlaschelli2007}.

\section*{Numerical Validation of Social Recommending}
To verify whether the proposed social recommending mechanism and
the network updating process can enhance the user experience, we
study how users' responses to the recommended items change over
time. In addition to user approval, we introduce a lower level
of user satisfaction by assuming that user $i$ says \emph{ok} to
item $\alpha$ if $r_{i\alpha}>0$. The ratios of the number of
approvals and ``okays'' to the total number of evaluations are
denoted by $p_a$ and $p_o$, respectively. When a given user $i$
evaluates item $\alpha$ with random attributes, the
average opinion is $\avg{r_{i\alpha}}=0$ and hence without
recommendation, $p_o=0.5$. Values of $p_o$ exceeding $0.5$
represent a working recommender system. As shown in Fig.~4, both
$p_o$ and $p_a$ increase quickly in the early stage of the
system's evolution and saturate at values considerably higher
than the initial ones.

We next check if the average quality of the evaluated items is
higher than it would be without recommendation. The intrinsic
quality of item $\alpha$ is defined as the sum of all the
elements of $\alpha$'s attribute vector,
$Q_{\alpha}:=\sum_{s=1}^{D}\mathbf{v}_{\alpha,s}/D$; the average
quality $\avg{Q}$ of all items is zero. We introduce the
effective average quality of evaluated items, $Q^*$, which is
weighted by the number of evaluations of each item. For example,
if an item with quality $-0.2$ was evaluated by $5$ users and
another item with quality $0.3$ was evaluated by $20$ users, the
corresponding value of $Q^*$ is
$(-0.2\times 5+0.3\times 20)/25=0.2$. A well-performing
recommender system should support spreading of high-quality
items and hence $Q^*$ should be high. As shown in Fig.~5, $Q^*$
increases in our system quickly from zero to approximately
$0.27$. Considering that the quality value of most items is
close to zero (less than $1\%$ of all items have quality greater
than the observed effective value $0.27$), this outcome is
creditable.

\section*{Conclusion and Discussion}
Uncovering common patterns of leader-follower networks is
important for our understanding of spreading processes in social
environments. We analyzed empirical data from four large-scale
real social networks where the notion of leadership can be
introduced and found indications of scale-free leadership
structures. We studied the social recommendation model inspired
by informal social recommending mechanisms (``word of mouth'')
that was studied in \cite{Medo2009}. We proposed a simplified
version of this model which was shown via agent-based
simulations to reproduce the observed power-law out-degree
distributions. The underlying mechanism leading to these
scale-free leadership structures can be summarized as ``good get
richer'': users with broad interests and good judgments are
likely to become popular leaders for the others. In our case,
broad interests are helpful to attract attention from the others
while good judgments ensure reliability of the received
recommendations. Although this result was obtained by a specific
recommendation model, its implications go beyond social
recommender systems. For example, the scale-free nature of
citation networks \cite{Price1965,Price1976,Redner1998,Redner2005,Perc2010a,Perc2010b} might be more fundamentally
explained by the present mechanism rather than by the notorious
``rich get richer'' mechanism \cite{Merton1968,Egghe1995,Barabasi1999}. The reason is that papers are cited by
scientists not only because they have already been cited many
times but mainly because they contain relevant and credible
results \cite{Price1976}. Notice that, the ``rich get richer" and ``good get richer" mechanisms are indeed related, depending on the criteria on goodness. For example, in evolutionary game, the criterion of a good player may be her/his cumulative wealth, and in scientific publications, the criterion of a good paper may be its cumulative citations. In such cases, the two mechanisms are not distinguishable. If only the network structure is observable, we can measure the strength of ``rich get richer" mechanism \cite{Jeong2003}, yet in principle we can say nothing about ``good get richer" mechanism. Additional information about each node's features, attributes, fitness and functionalities may drive us to more in-depth understanding about the existence of ``good get richer" mechanism. In this point of view, the ``good get richer" mechanism can be considered as a deeper mechanism underlying the observed ``rich get richer" phenomenon.

Furthermore, our agent-based simulations reveal that the
proposed model is an effective tool for quality information
filtering and it is also efficient in requiring very little
computation. These noticeable features are of particular
relevance for resource-sharing services which are recently
experiencing increasing popularity. Most of them (take
\emph{digg.com}, \emph{reddit.com} and \emph{wikio.com} as
examples) still adopt the traditional organization in which
resources are ranked by popularity and divided into categories
created by a top-down approach. Known recommendation techniques
are also designed in a centralized way where the systems, rather
than the users, decide what to recommended to whom
\cite{Billsus2007}. By contrast, systems like
\emph{delicious.com} and \emph{twitter.com} have implemented the
possibility to recommend and to have something recommended
by other users. The fast growth of these online communities
\cite{Kumar2006} as well as the fact that users prefer
recommendations coming from their social circle \cite{Sinha2001,Huang2010}
make social recommendation a promising way to better organize
and deliver online resources and to enhance online social
contacts. While we neglected some relevant social factors like
friendship and reciprocity and could not provide analytical
solution of the proposed model, this paper offers various
insights to the dynamics of resource-sharing systems and
provides a starting point for their future studies.

%This work is partially supported by the Swiss National Science
%Foundation under Grant No. 200020-121848, and the National Natural
%Science Foundation of China under Grant Nos. 10635040, 90924011 and
%60973069.

\newpage

\begin{table}
\centering
\caption{Basic characteristics and results of statistical
analysis for the studied leadership networks. $N$ represents the
number of users, $E$ represents the number of links,
$x_{\mathrm{min}}$ is the lower bound of the range fit by a
power-law distribution, $\alpha$ is the corresponding power-law
exponent obtained by maximum likelihood estimation and $KS$ is
the goodness-of-fit value based on the Kolmogorov-Smirnov
statistic \cite{Clauset2009}.}
\begin{tabular}{lrrrrr}
\toprule
Dataset   &        $N$ &           $E$ & $x_{\mathrm{min}}$ & $\alpha$ & $KS$  \\
\midrule
Delicious &    392,251 &     1,686,131 &                 20 &     2.82 & 0.010\\
Flickr    &  1,441,432 &    22,613,981 &                 10 &     1.78 & 0.021\\
Twitter   & 35,689,148 & 1,468,365,183 &                 50 &     1.88 & 0.033\\
YouTbue   &    570,774 &     4,945,382 &                 10 &     2.13 & 0.013\\
\bottomrule
\end{tabular}
\end{table}

\begin{figure}
\centering
\includegraphics[scale=0.27]{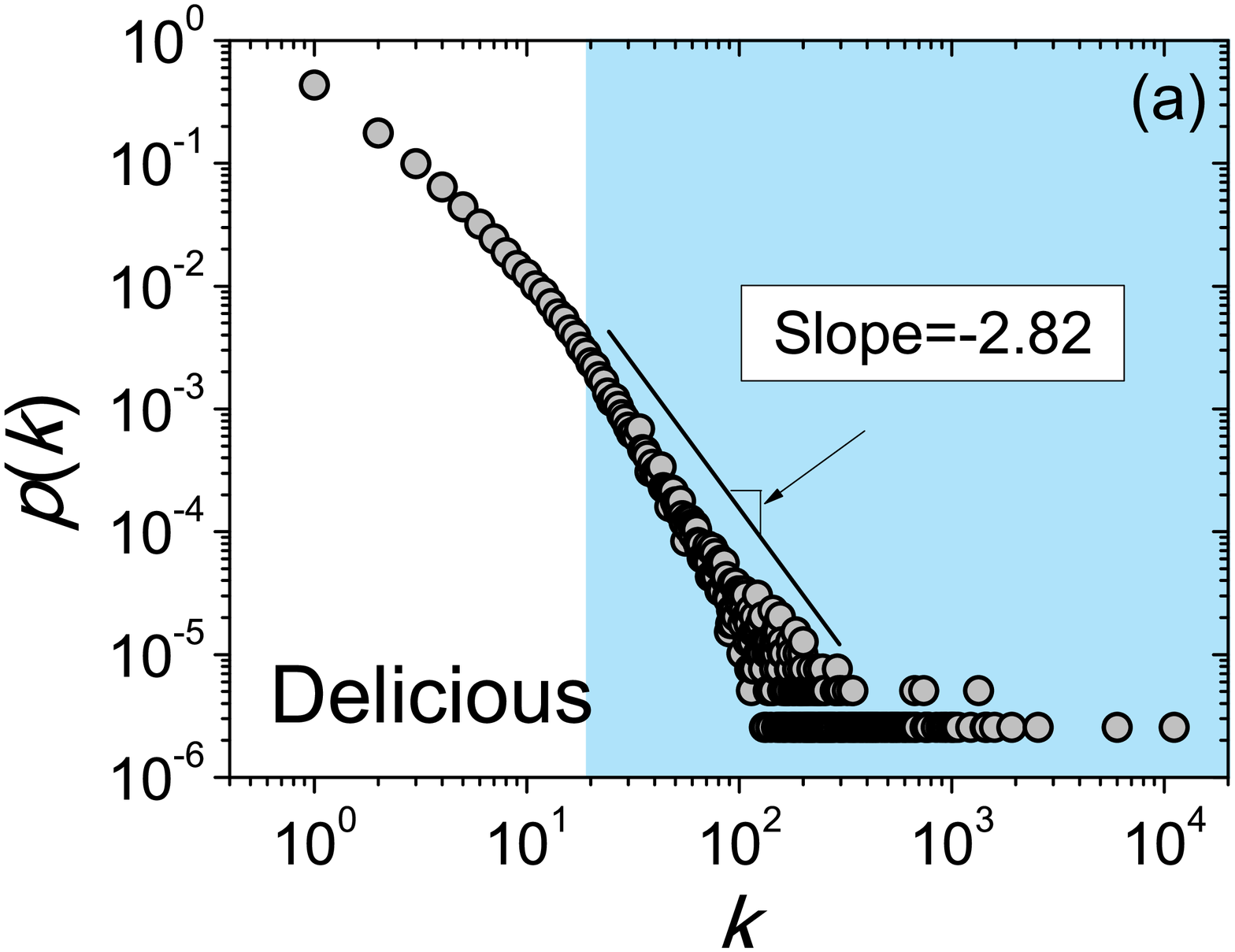}
\includegraphics[scale=0.27]{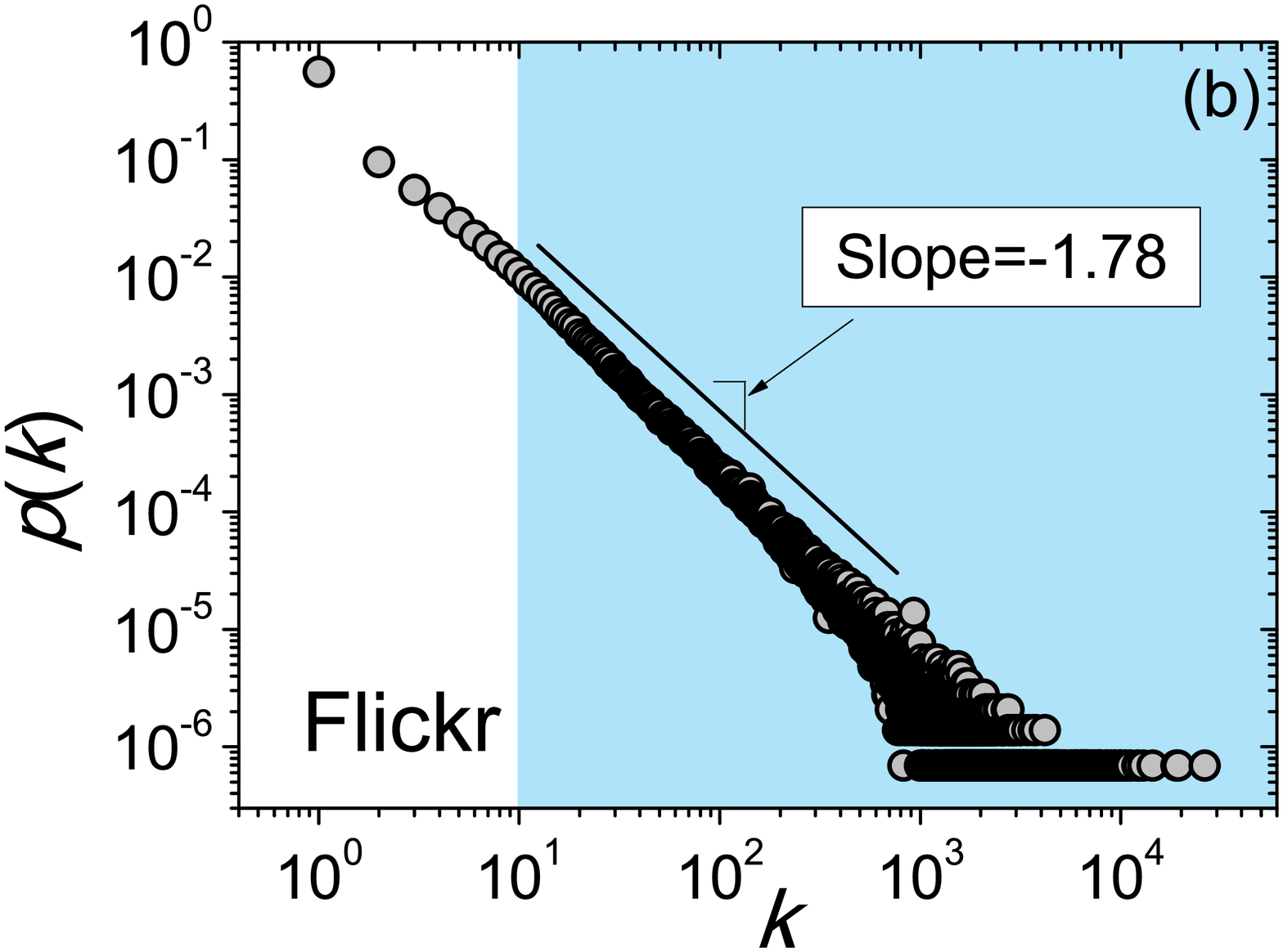}
\includegraphics[scale=0.27]{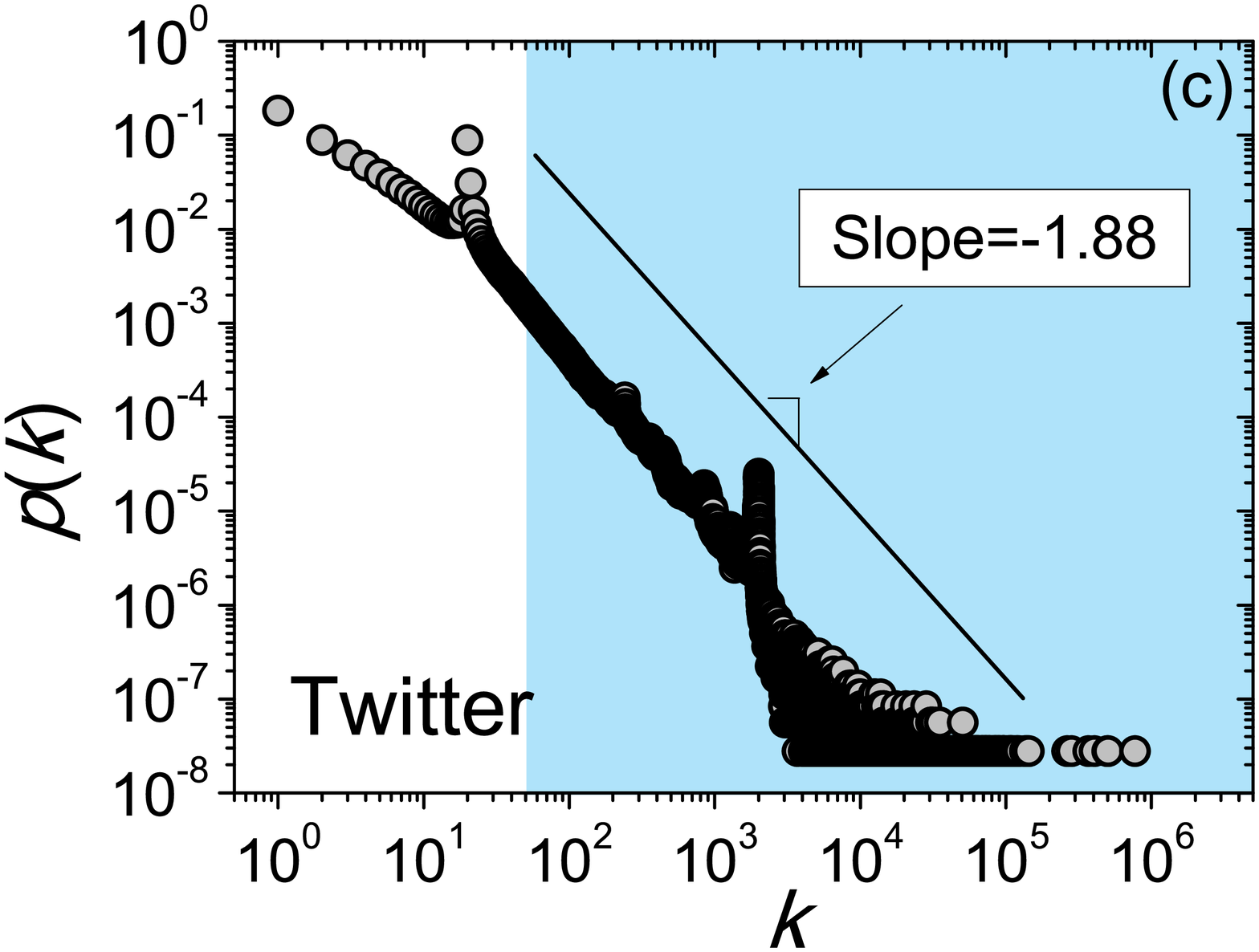}
\includegraphics[scale=0.27]{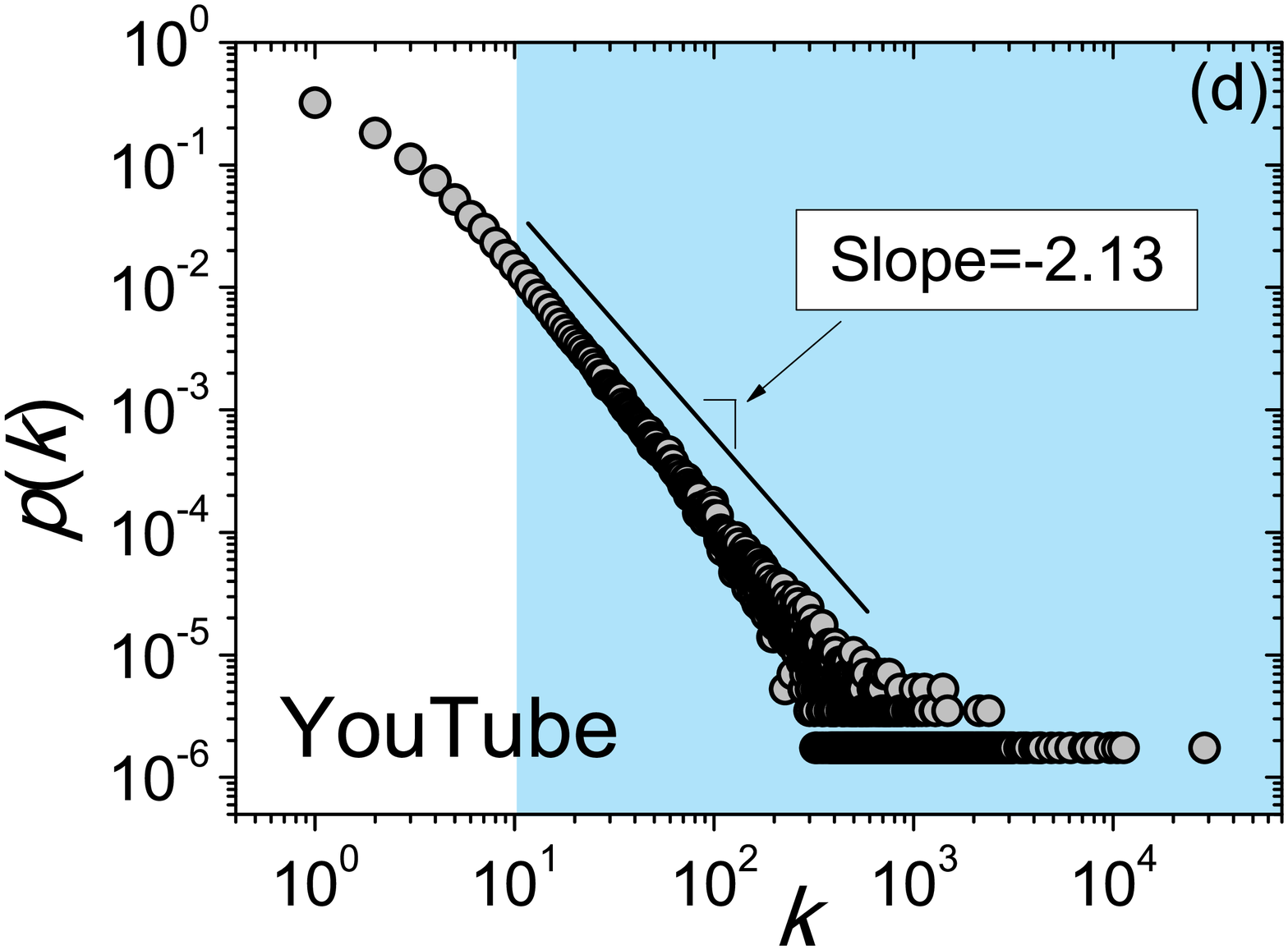}
\caption{{\bf Scale-free leadership structure -- empirical results.} Out-degree distributions of the studied leadership
networks and their power-law fits. Shaded areas in the figures
show the range where the data is best described by a power-law
distribution (they are delimited by $x_{\mathrm{min}}$
minimizing the $KS$ statistic).}
\end{figure}

\begin{figure}
\centering
\includegraphics[scale=0.5]{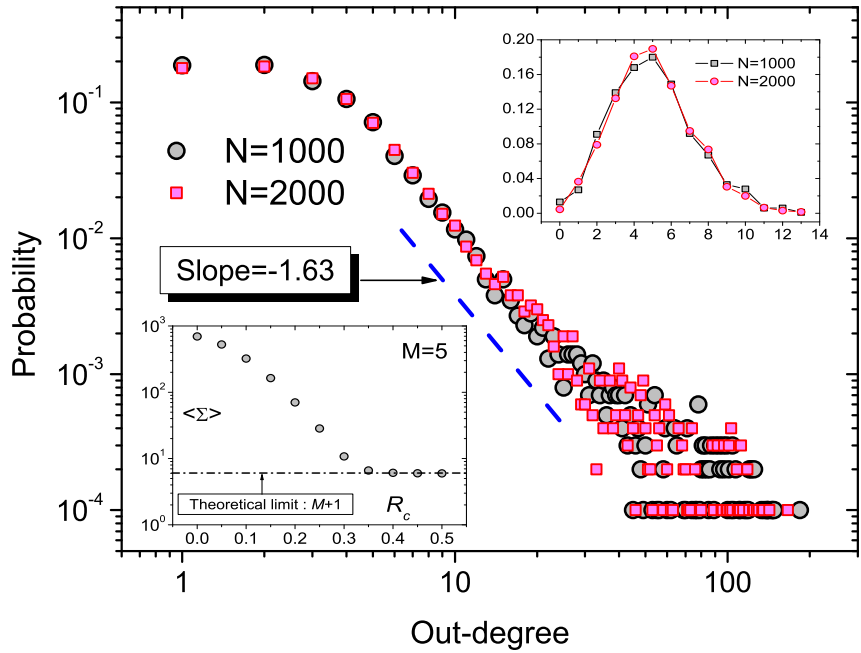}
\caption{{\bf Scale-free leadership structure -- simulation results.} Out-degree distributions of the
resulting leadership networks at time step $10^6$ for $M=5$,
$T=100$, $D=13$, $R_c=0.2$ and different values of $N$. The
upper-right inset displays the initial out-degree distributions.
The lower-left inset shows the average number of readers of an
item as a function of $R_c$ for $N=1\,000$. The thick dashed
line with slope $-1.6$ is shown as a guide to the eye. All data
points reported here and later are averaged over 10
realizations.}
\end{figure}

\begin{figure}
\centering
\includegraphics[scale=0.27]{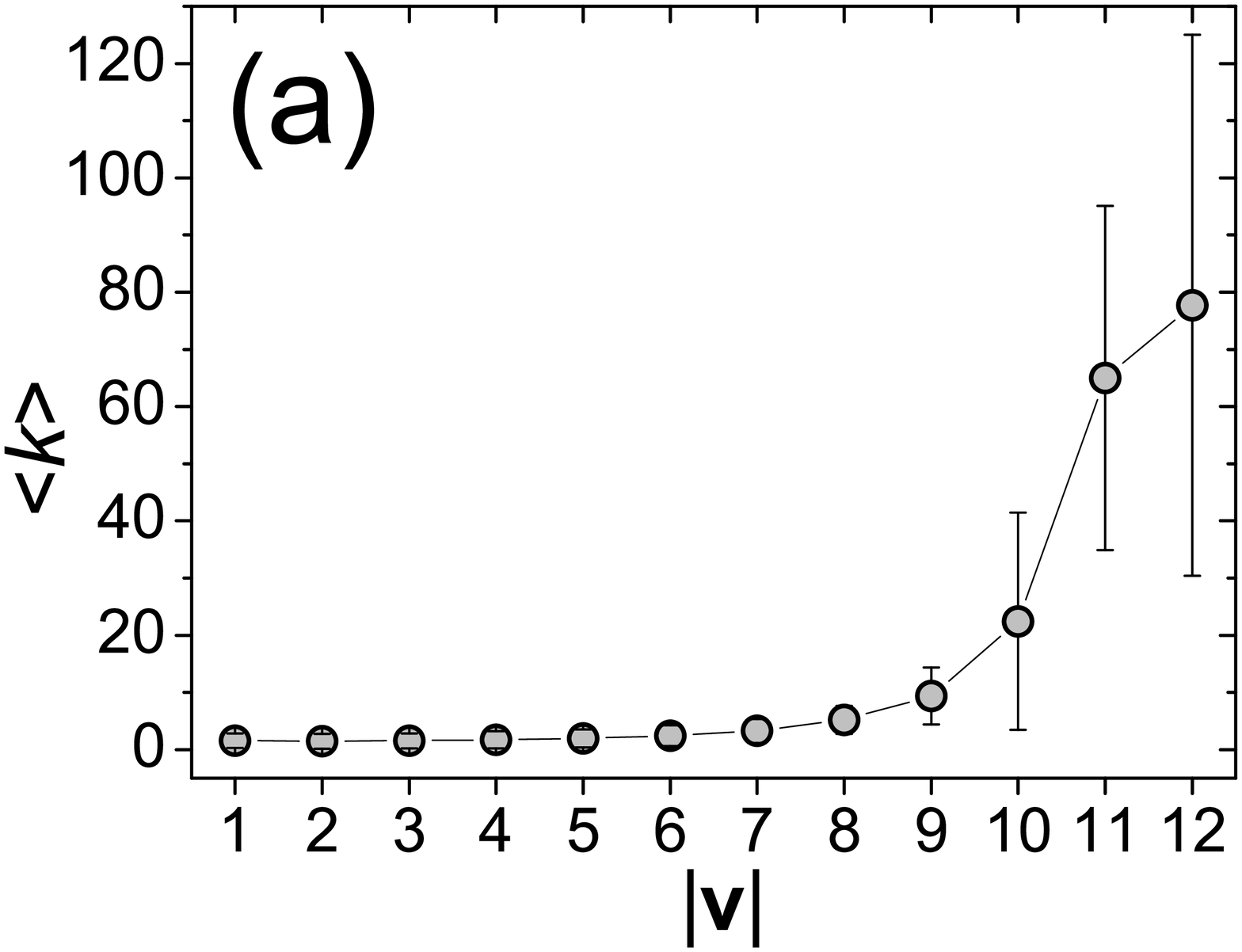}
\includegraphics[scale=0.27]{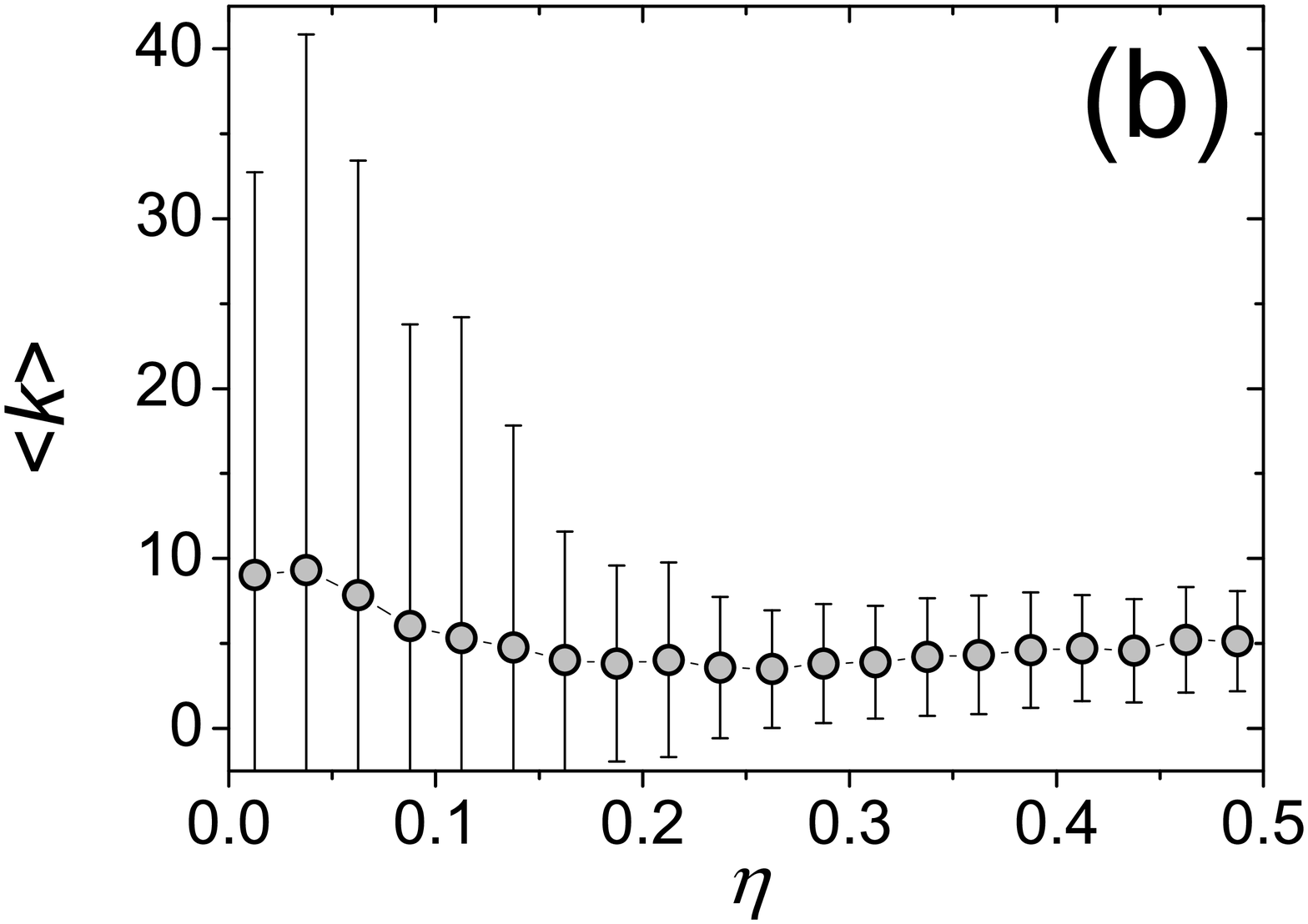}
\includegraphics[scale=0.27]{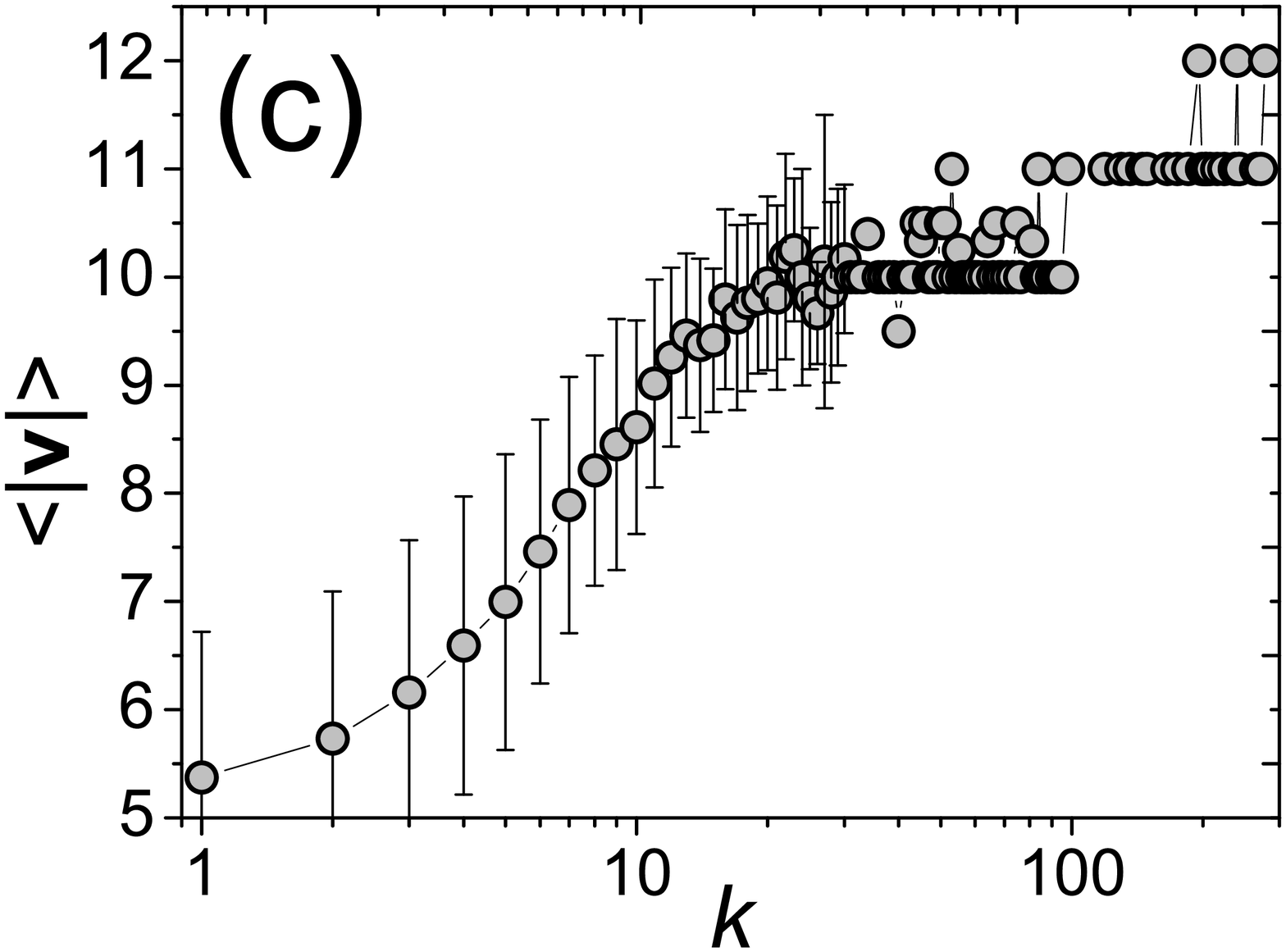}
\includegraphics[scale=0.27]{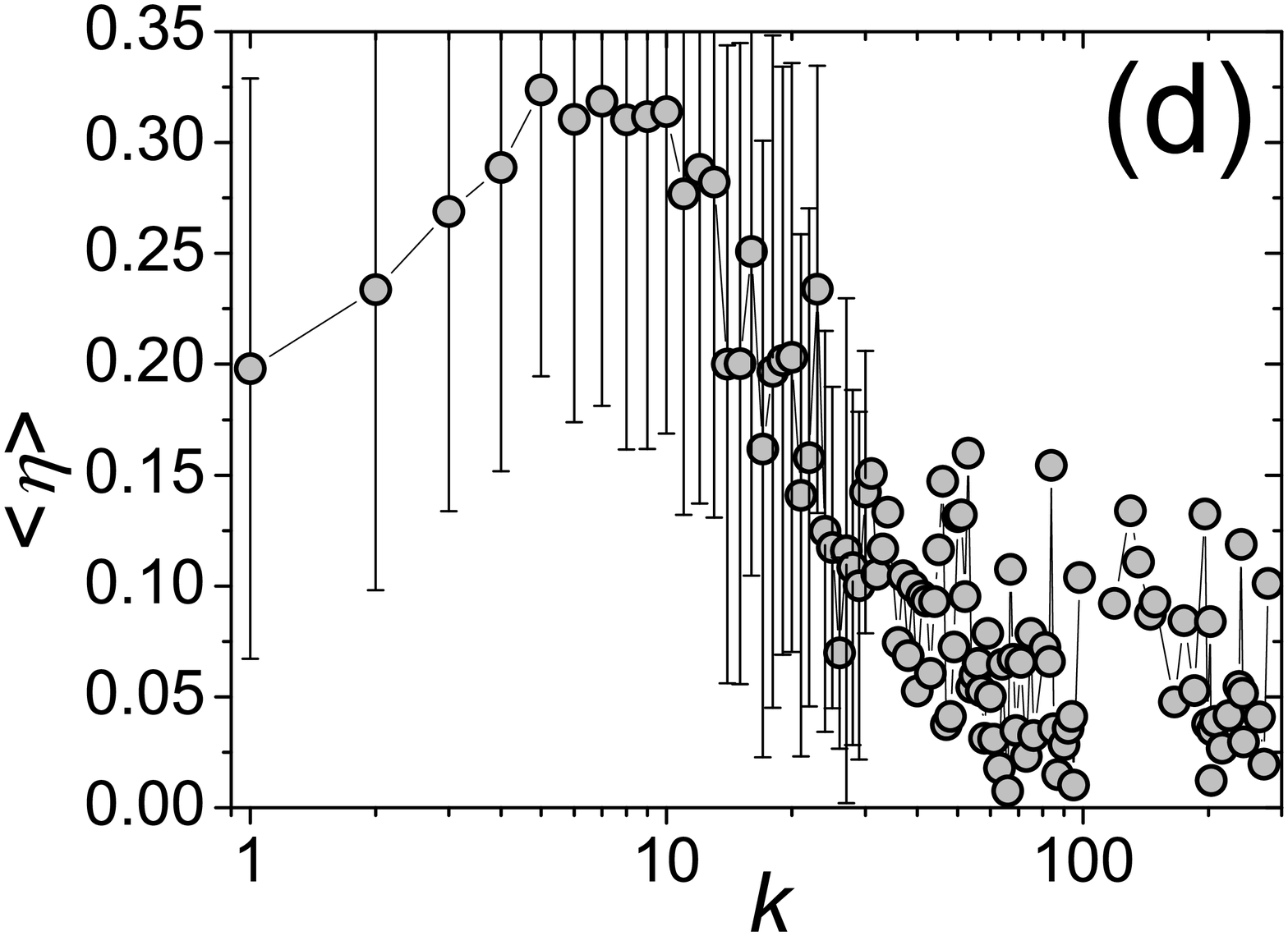}
\caption{{\bf Broad interests and good judgments make a leader.} Dependencies between the leadership strength and the
scope of interests (a,c), and the quality of evaluations (b,d),
respectively. The data points and error bars correspond to mean
values and standard deviations. In (c) and (d), when $k>30$,
there is not enough data to obtain credible error bars, hence
they are not shown. The population size is $N=1000$; other
parameter values are the same as in Fig.~2.}
\end{figure}

\begin{figure}
\centering
\includegraphics[scale=0.27]{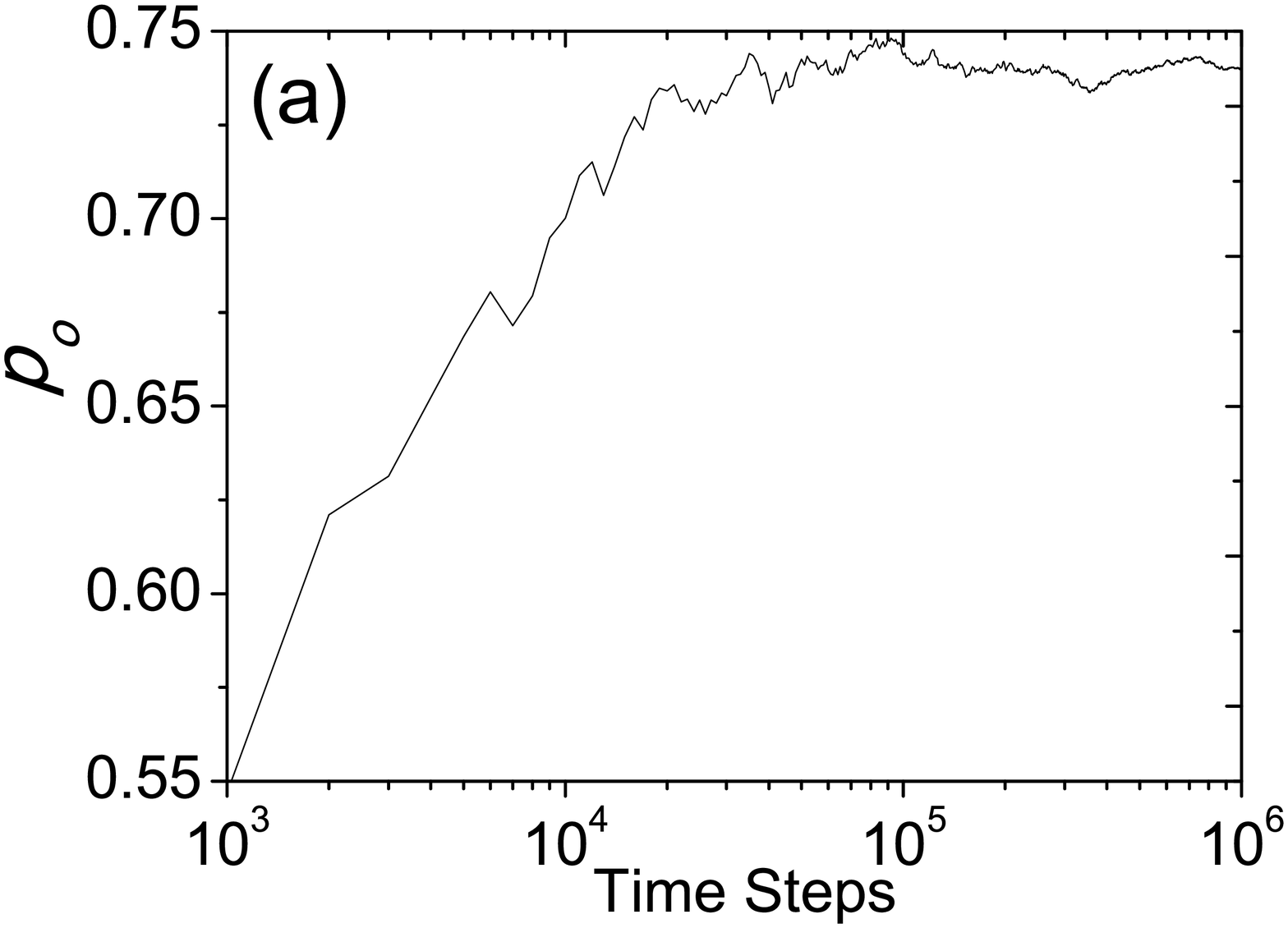}
\includegraphics[scale=0.27]{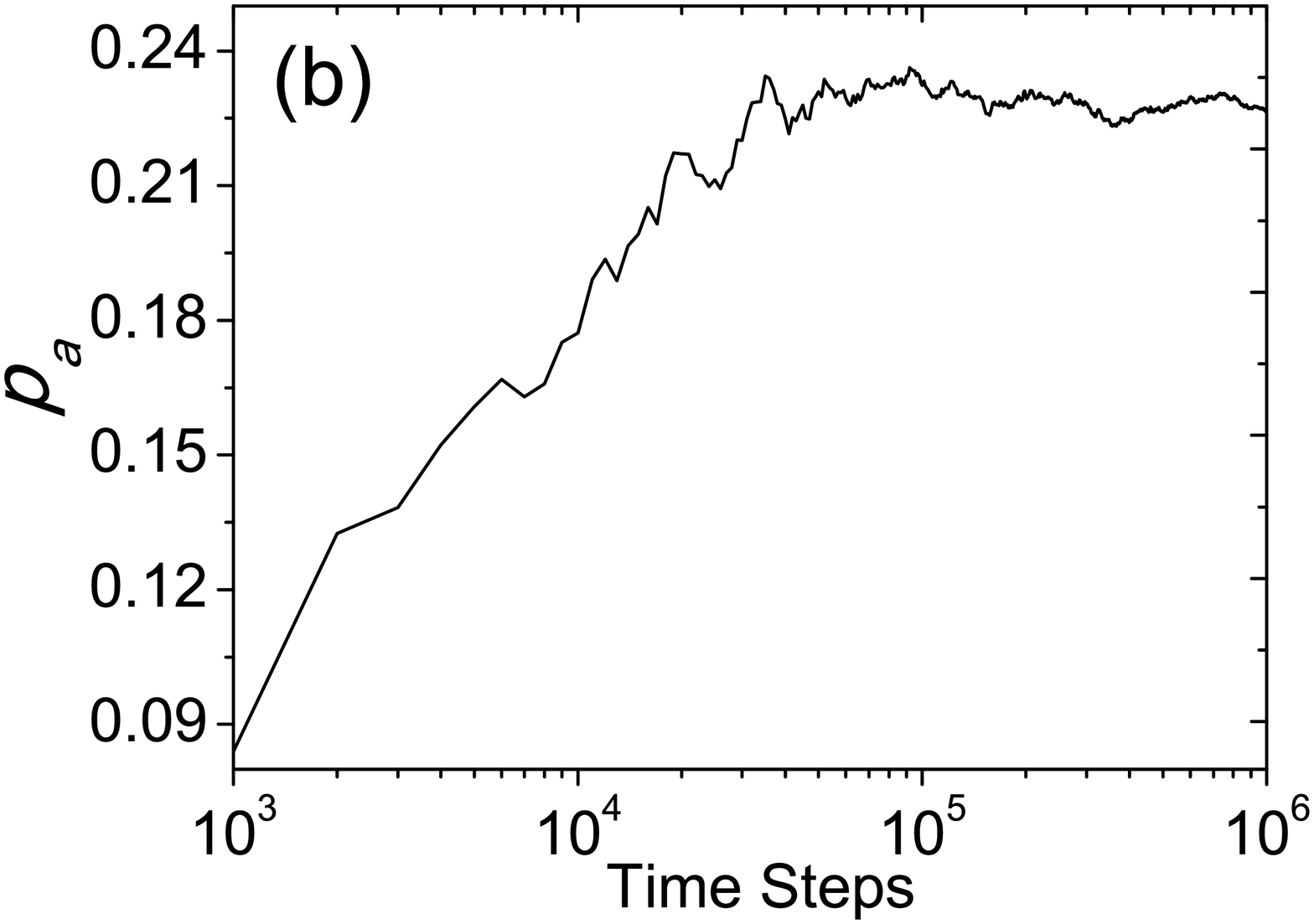}
\caption{{\bf User experience is enhanced by this social recommender system.} Probabilities of saying \emph{ok} (a) and approving
(b) items versus time. Values shown at time $t$ correspond to
the average $p_o$ and $p_a$ in time steps from $t-10^3$ to $t$.
Parameter values are the same as in Fig.~3.}
\end{figure}

\begin{figure}
\centering
\includegraphics[scale=0.3]{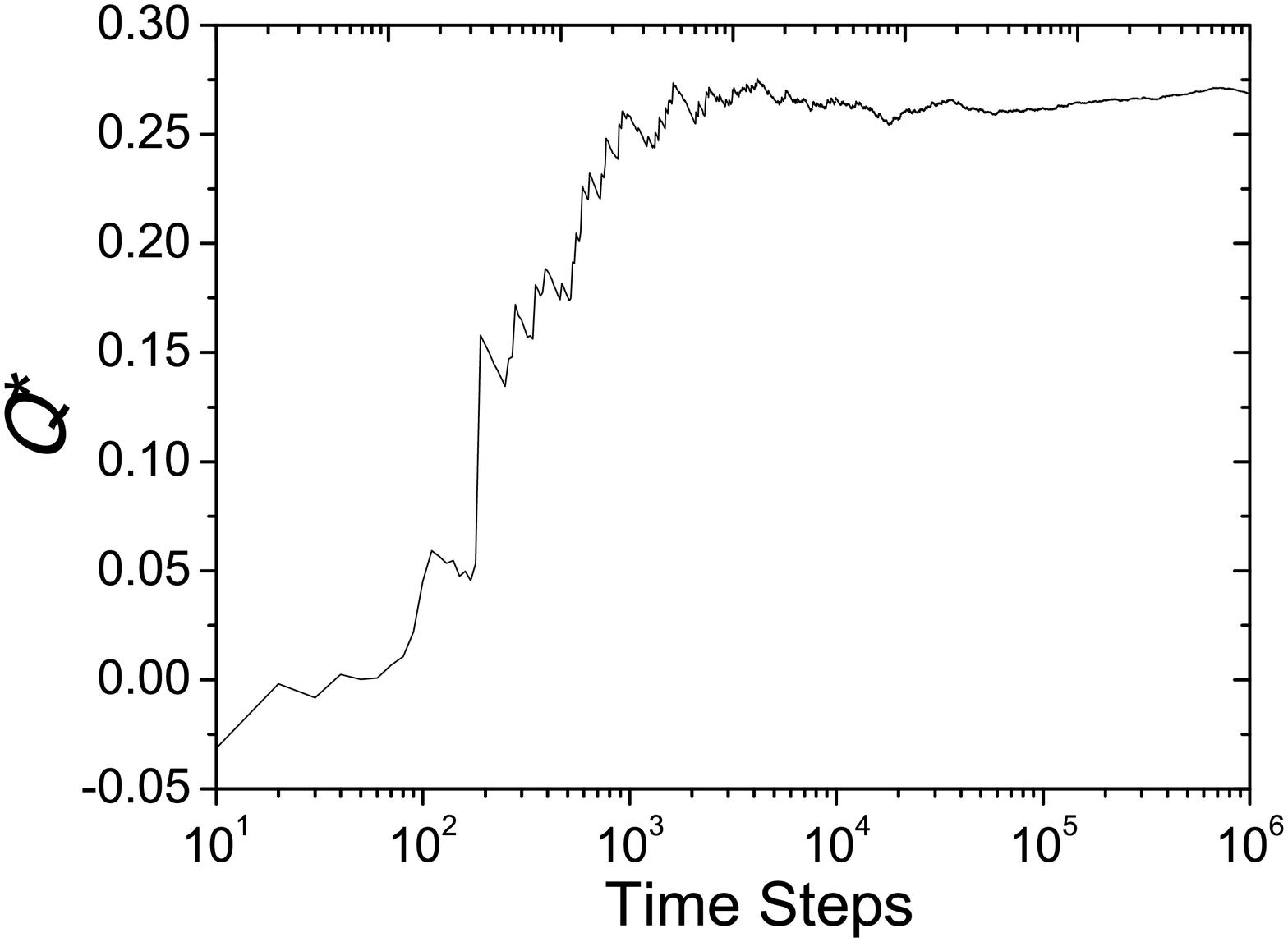}
\caption{{\bf Good news will live longer while bad news will die out soon.} Time evolution of the effective quality $Q^*$ of
evaluated items. Parameter values are the same as in Fig.~3.}
\end{figure}


\begin{thebibliography}{99}
\bibitem{Watts2007} Watts DJ (2007) A twenty-first century science. \emph{Nature} 445: 489.
\bibitem{Borgatti2009} Borgatti SP, Mehra A, Brass DJ, Labianca G (2009) Network analysis in the social sciences. \emph{Science}, 323: 892--895.
\bibitem{Anghel2004} Anghel M, Toroczkai Z, Bassler KE, Korniss G (2004) Competition-driven network dynamics: Emergence of a scale-free leadership structure and collective efficiency. \emph{Phys Rev Lett} 92: 058701.
\bibitem{Lo2005} Lo TS, Chan KP, Hui P-M, Johnson NF (2005) Theory of enhanced performance emerging in a sparsely connected competitive population. \emph{Phys Rev E} 71: 050101.
\bibitem{Lee2006} Lee SH, Jeong H (2006) Effects of substrate network topologies on competition dynamics. \emph{Phys Rev E} 74: 026118.
\bibitem{Savarimuthu2008} Savarimuthu BTR, Cranefield S, Purvis M, Purvis M (2008) Role model based mechanism for norm emergence in artifical agent societies. \emph{Lect Notes Comput Sci} 4870: 203--217.
\bibitem{Szolnoki2008} Szolnoki A, Perc M, Danku Z (2008) Making new connections towards cooperation in the prisoner's dilemma game. \emph{EPL} 84: 50007.
\bibitem{Szolnoki2009a} Szolnoki A, Perc M (2009) Resolving social dilemmas on evolving random networks. \emph{EPL} 86: 30007.
\bibitem{Szolnoki2009b} Szolnoki A, Perc M (2009) Emergence of multilevel selection in the prisoner's dilemma game on coevolving random networks. \emph{New J Phys} 11: 093033.
\bibitem{Poncela2008} Poncela J, G\'omez-Garde\"nes J, Flor\'ia LM, S\'anchez A, Morena Y (2008) Complex Cooperative Networks from Evolutionary Preferential Attachment. \emph{PLoS ONE} 3(6): e2449.
\bibitem{Resnick1997} Resnick P, Varian HR (1997) Recommender systems. \emph{Commun ACM}  40: 56--58.
\bibitem{Herlocker2004} Herlocker JL, Konstan JA, Terveen K, Riedl JT (2004) Evaluating Collaborative Filtering Recommender Systems. \emph{ACM Trans Inf Syst} 22: 5--53.
\bibitem{Schafer2007} Schafer J, Frankowski D, Herlocker J, Sen S (2007) Collaborative filtering recommender systems. \emph{Lect Notes Comput Sci} 4321: 291--324.
\bibitem{Maslov2001} Maslov S, Zhang Y-C (2001) Extracting hidden information from knowledge networks. \emph{Phys Rev Lett} 87: 248701.
\bibitem{gravity2007} Tak\'acs G, Pil\'aszy I, N\'emeth B, Tikk D (2007) Major components of the gravity recommendation system. \emph{ACM SIGKDD Explorations Newsletter} 9: 80--83.
\bibitem{Ren2008} Ren J, Zhou T, Zhang Y-C (2008) Information filtering via self-consistent refinement. \emph{EPL} 82: 58007.
\bibitem{Zhang2007} Zhang Y-C, Blattner M, Yu Y-K (2007) Heat conduction process on community networks as a recommendation model. \emph{Phys Rev Lett} 99: 154301.
\bibitem{Zhou2007} Zhou T, Ren J, Medo M, Zhang Y-C (2007) Bipartite network projection and personal recommendation. \emph{Phys Rev E} 76: 046115.
\bibitem{Zhou2010} Zhou T, Kuscsik Z, Liu J-G, Matus M, Wakeling JR, Zhang Y-C (2010) Solving the apparent diversity-accuracy dilemma of recommender systems. \emph{Proc Natl Acad Sci USA} 107: 4511--4514.
\bibitem{Ziegler2004} Ziegler C-N, Lausen G (2004) Paradigms for Decentralized Social Filtering Exploiting Trust Network Structure. \emph{Lect Notes Comput Sci} 3291: 840--858.
\bibitem{Bonhard2006} Bonhard P, Sasse MA (2006) 'Knowing me, knowing you'---Using profiles and social networking to improve recommender systems. \emph{BT Technology J} 24: 84--98.
\bibitem{Sinha2001} Sinha R, Swearingen K (2001) Comparing Recommendations made by Online Systems and Friends. Proc DELOS-NSF Workshop on Personalization and Recommender Systems in Digital Libraries \emph(Dublin, Ireland).
\bibitem{Huang2010} Huang J, Cheng X-Q, Guo J, Shen H-W, Yang K (2010) Social Recommendation with Interpersonal Influence. Proc 19th Euro Conf Artif Int \emph(Dublin, Ireland), p 601--606.
\bibitem{Golbeck2008} Golbeck J (2008) Weaving a Web of Trust. \emph{Science} 321: 1640--1641.
\bibitem{Hammond2005} Hammond T, Hannay T, Lund B, Scott J (2005) Social bookmarking tools (I): a general review. \emph{D-Lib Magazine} 11(4).
\bibitem{Chen2008} Chen H-Q, Cheng X-Q, Liu Y (2008) Finding Core Members in Virtual Communities. Proc 17th Intl Conf WWW \emph{ACM Press, New York} 1233 p.
\bibitem{Medo2009} Medo M, Zhang Y-C, Zhou T (2009) Adaptive model for recommendation of news. \emph{EPL} 88: 38005.
\bibitem{Cimini2011} Cimini G, Medo M, Zhou T, Wei D, Zhang Y-C (2011) Heterogeneity, quality, and reputation in an adaptive recommendation model. \emph{Eur Phys J B} 80: 201--208.
\bibitem{Wei2011} Wei D, Zhou T, Cimini G, Wu P, Liu W, Zhang Y-C (2011) Effective mechanism for social recommendation of news. \emph{Physica A} 390: 2117--2126.
\bibitem{Pastor2003} Pastor-Satorras R, Vespignani A (2003) Epidemics and Immunization in Scale-free Networks in Handbook of Graph and Networks, edited by Bornholdt S and Schuster HG (Wiley-VCH, Berlin).
\bibitem{Zhou2006} Zhou T, Fu Z-Q, Wang B-H (2006) Epidemic dynamics on complex networks. \emph{Prog Nat Sci} 16: 452--463.
\bibitem{ZhangZK2010} Zhang Z-K, Liu C (2010) A hypergraph model of social tagging networks. \emph{J Stat Mech} 2010: P10005.
\bibitem{Goldstein2004} Goldstein ML, Morris SA, Yen GG (2004) Problems with fitting to the power-law distribution. \emph{Eur Phys J B} 41: 255--258.
\bibitem{Clauset2009} Clauset A, Shalizi CR, Newman MEJ (2009) Power-law distributions in empirical data \emph{SIAM Rev} 51: 661--703.
\bibitem{Radicchi2009} Radicchi F, Fortunato S, Markines B, Vespignani A (2009) Diffusion of scientific credits and the ranking of scientists. \emph{Phys Rev E} 80: 056103.
\bibitem{Mislove2007} Mislove A, Marcon M, Gummadi KP, Druschel P, Bhattacharjee B (2007) Measurement and analysis of online social networks. Proc 7th ACM SIGCOMM Conf on Internet Measurement \emph{ACM Press, New York} 29 p.
\bibitem{Kwak2010} Kwak G, Lee C, Park H, Moon S (2010) What is Twitter, a social network or a news media? Proc 19th WWW \emph{ACM Press, New York} 591 p.
\bibitem{Lee2010} Lee SH, Kim P-J, Ahn Y-Y, Jeong H (2010) Googling social interactions: Web search engine based social network construction. \emph{PLoS ONE} 5: e11233.
\bibitem{Lu2011} L\"{u} L, Zhang Y-C, Yeung CH, Zhou T. Leaders in Social Networks: the Delicious Case. \emph{PLoS ONE} (to be published).
\bibitem{Caldarelli2007} Caldarelli G (2007) Scale-Free Networks. \emph{Oxford Press, New York}.
\bibitem{Mitzenmacher2003} Mitzenmacher M (2003) A Brief History of Generative Models for Power Law and Lognormal Distributions. \emph{Internet Mathematics} 1: 226--251.
\bibitem{Merton1968} Merton RK (1968) The Matthew Effect in Science. \emph{Science} 159: 56-63.
\bibitem{Egghe1995} Egghe L, Rousseau R (1995) Generalized success-breeds-success principle leading to time-dependent informetric distributions. \emph{J Am Soc Inf Sci} 46: 426-445.
\bibitem{Barabasi1999} Barab\'asi A-L, Albert R (1999) Emergence of scaling in random networks. \emph{Science} 286: 509--513.
\bibitem{Valverde2002} Valverde S, Cancho RF, Sol\'e RV (2002) Scale-Free Networks from Optimal Design. \emph{EPL} 60: 512--517.
\bibitem{Baiesi2003} Baiesi M, Manna SS (2003) Scale-free networks from a Hamiltonian dynamics. \emph{Phys Rev E} 68: 047103.
\bibitem{Kim2005} Kim BJ, Trusina A, Minnhagen P, Sneppen K (2005) Self Organized Scale-Free Networks from Merging and Regeneration. \emph{Eur Phys J B} 43: 369--372.
\bibitem{Perotti2009} Perotti JI, Billoni OV, Tamarit FA, Chialvo DR, Cannas SA (2009) Emergent Self-Organized Complex Network Topology out of Stability Constraints. \emph{Phys Rev Lett} 103: 108701.
\bibitem{Caldarelli2002} Caldarelli G, Capocci A, De Los Rios P, Mu\~noz MA (2002) Scale-Free Networks from Varying Vertex Intrinsic Fitness. \emph{Phys Rev Lett} 89: 258702.
\bibitem{Garlaschelli2007} Garlaschelli D, Capocci A, Caldarelli G (2007) Self-organized network evolution coupled to extremal dynamics. \emph{Nat Phys} 3: 813--817.
\bibitem{Price1965} Price DJ de S (1965) Networks of Scientific Papers. \emph{Science} 149: 510--515.
\bibitem{Price1976} Price DJ de S (1976) A general theory of bibliometric and other cumulative advantage processes. \emph{J Am Soc Inf Sci} 27: 292--306.
\bibitem{Redner1998} Redner S (1998) How popular is your paper? An empirical study of the citation distribution. \emph{Eur Phys J B} 4: 131--134.
\bibitem{Redner2005} Redner S (2005) Citation Statistics from 110 Years of Physical Review. \emph{Physics Today} 58(6): 49--54.
\bibitem{Perc2010a} Perc M (2010) Zipf's law and log-normal distributions in measures of scientific output across fields and institutions: 40 years of Slovenia's research as an example. \emph{J Informetrics} 4: 358-364.
\bibitem{Perc2010b} Perc M (2010)  Growth and structure of Slovenia's scientific collaboration network \emph{J Informetrics} 4: 475-482.
\bibitem{Jeong2003} Jeong H, Nda Z, Barab\'asi A-L (2003) Measuring preferential attachment in evolving networks. \emph{Europhys Lett} 61: 567-572.
\bibitem{Billsus2007} Billsus D, Pazzani MJ (2007) Adaptive news access. \emph{Lect Notes Comput Sci} 4321: 550--570.
\bibitem{Kumar2006} Kumar R, Novak J, Tomkins A (2006) 	Structure and evolution of online social networks. Proc 12th ACM SIGKDD \emph{ACM Press, New York} 611 p.
\end{thebibliography}
\end{document}